\documentclass[twocolumn,twoside,slac_two]{revtex4}
\usepackage{graphicx}
\usepackage{fancyhdr}
\begin{document}

\title{CLEO-c Hot Topics}

%

\author{R. Poling}
\affiliation{University of Minnesota, 116 Church Street SE, 
Minneapolis, MN 55455}

\begin{abstract}
Selected recent results and future prospects for the CLEO-c experiment 
at CESR are reviewed.  The topics covered include measurements of leptonic 
and semileptonic charm decays made with data collected at the $\psi(3770)$ 
resonance and results from a scan of the center-of-mass energy 
range from 3970 to 4260 MeV addressing the details of open-charm 
production and properties of the $Y(4260)$ state observed last year 
by BaBar.  
\end{abstract}

\maketitle

\thispagestyle{fancy}

\section{Introduction}
\label{sec:intro}

The CLEO-c experiment at the Cornell Electron Storage Ring (CESR) has 
provided our field with an unanticipated opportunity to turn the power of
a state-of-the-art magnetic spectrometer developed for studying $B$ decays to long-neglected
problems of charm physics \cite{Briere:2001rn}.  This paper is the first of several from CLEO-c 
in these proceedings, and includes a number of interesting early results from the CLEO-c charm 
program.  Additional details and other topics are covered elsewhere in this volume in 
the papers of Cinabro, Stone, Vogel, and Wiss.

The remainder of this introduction provides an overview of the accelerator, detector, 
and motivation for CLEO-c.  Sect.~\ref{sec:lep} describes results in leptonic and 
semileptonic $D$ decays.  Sect.~\ref{sec:scan} describes results on closed- and open-charm 
production in the center-of-mass energy range from 3970 to 4260 MeV from a scan completed 
in fall 2005.  In Sect.~\ref{sec:scan} I provide a short summary and projection of what will come 
from CLEO-c before it concludes operation in 2008.

The CESR-c/CLEO-c project arose from the recognition that we had available to us a facility of
great versatility that was no longer competitive for the job it was designed to do.  The asymmetric
B factories came on line with such excellent performance that it quickly became apparent that CESR and
CLEO~III could not perform at a comparable level.  We recognized that we needed 
a niche that is complementary to Belle and BaBar.  Plans were formulated in coordination with the 
CESR accelerator group to implement operations at about one-third of CESR's design 
energy, with charm physics as the principal objective.  While conversion plans were being executed, 
we conducted productive runs on the narrow bottomonium resonances and made 
initial tests of $\psi(2S)$ operations.

By early 2004 the transformation was complete, and we had implemented Eq.~\ref{eq-CESR}:
\begin{equation}
{\rm CESR-c} \simeq {\rm CESR} - 6.5~{\rm GeV} + 12({\rm wigglers}),
\label{eq-CESR}
\end{equation}
where the superconducting wiggler magnets provide enhanced damping of the stored beams that 
is required because of
the reduction of CESR's energy from $\sim 10$~GeV to $\sim 4$~GeV.  
The wigglers functioned well, but luminosity performance has fallen short
of expectations.  Various problems have been uncovered, and many improvements implemented,
including enhanced instrumentation in the form of a fast-feedback luminosity monitor.  With
more detailed measurements and an extensive simulation program, it was determined that
specific-luminosity limitations were partially due to energy sensitivity in the compensation 
for CLEO's solenoidal magnetic field.  This led to several optics and operational changes for
CESR, in particular a new ``antisolenoid'' magnet modeled on a similar compensation scheme devised 
at Frascati.  It was installed in January, 2006 and the 
benefits when running resumed were immediate, with an initial $\sim 20$\% improvement in instantaneous luminosity 
(to $6.7 \times 10^{31}$~cm$^{-2}$s$^{-1}$) and in ``best-day'' integrated luminosity 
(to $\sim 4.2$~pb$^{-1}$).
Machine studies aimed at achieving the full benefit of the antisolenoid and 
other improvements 
continue, and we expect to collect at least 5~pb$^{-1}$ per day in coming runs.

\begin{figure}[b]
\centering
\includegraphics[width=80mm]{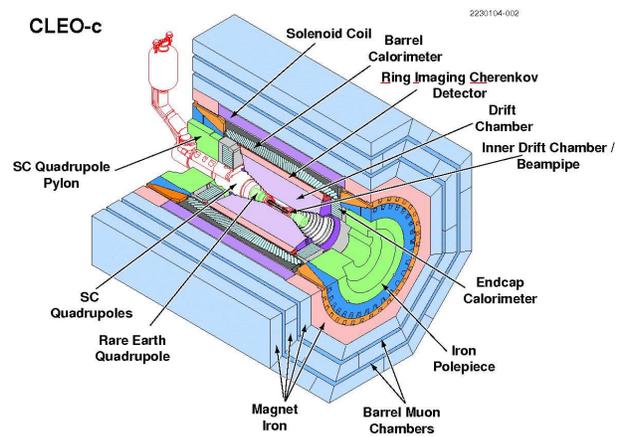}
\caption{CLEO-c detector.} \label{fig:CLEOc_detector}
\end{figure}

In preparing for the new energy regime, only modest modifications were needed for CLEO's detector 
(Fig.~\ref{fig:CLEOc_detector}), 
as described in Eq.~\ref{eq-CLEO}:
\begin{equation}
{\rm CLEO-c} \simeq {\rm CLEO~III} - {\rm SVX} + {\rm ZD} - 0.5~{\rm T},
\label{eq-CLEO}
\end{equation}
where the previous silicon vertex detector (SVX) was replaced by a low-mass all-stereo drift chamber (ZD)
and the magnetic field was reduced to mitigate the effect of soft particles curling up in the tracking
volume.  The conversion to CLEO-c has been a great success, and the detector remains at the top of its
game, with charged-particle momentum resolution ($\Delta p/p = 0.6\%$ at 1~GeV/$c$) and electromagnetic
calorimetry ($\Delta E/E$ = 2.2\% at 1~GeV and 5\% at 100~MeV) superior to other detectors that have operated
near charm threshold.  Excellent electron identification and hadron identification (RICH and $dE/dx$) 
are likewise unprecedented tools for charm physics in this energy range.

The opportunity that CLEO-c represented at the time of its proposal is now being realized, in
spite of the reduction by roughly a factor of 4 in expected integrated luminosity.  Impact on the worldwide
CKM program is both direct and indirect, and will be leveraged powerfully in the interpretation of $B$ decays
through precision tests of theoretical tools like lattice QCD.  In addition, ``engineering'' input such as
branching fractions for key normalization modes in $D$ and $D_s$ decays will be crucial.

The data samples that have been used for the results presented 
in this paper include a total of 281~pb$^{-1}$ collected
at the $\psi(3770)$ resonance.  Many initial results from the first 
57~pb$^{-1}$ of this sample have been published, and 
initial results from the full sample have begun to appear, 
including several that are new for FPCP 2006.  The $D_s$ scan run 
consisted of a total of about 60~pb$^{-1}$ at twelve energy points above the
$\psi(3770)$, and it immediately led to CLEO-c's first dedicated $D_s$ run at 4170~MeV, 
which was completed in April, 2006.  There has also been a small amount of data collected
at the $\psi(2S)$ resonance, split equally between CLEO~III and CLEO-c.  While totaling only 
5.6~pb$^{-1}$, this sample has produced 40 analysis results,
some of which are described elsewhere in these proceedings.

\section{\boldmath$D$ Physics at \boldmath$\psi(3770)$: Leptonic and Semileptonic}
\label{sec:lep}

Much of the progress in heavy-flavor physics has come through detailed studies of leptonic and semileptonic 
decays.  In $B$ physics we use events with leptons to gain access to unknown Standard Model parameters with
less uncertainty due to strong-interaction effects than in purely hadronic decays.  As experimental precision
has improved, however, essentially every determination of CKM parameters, especially of the magnitudes $|V_{cb}|$
and $|V_{ub}|$, has become systematics-limited by theoretical uncertainties.  Studies of charm semileptonic
and leptonic decays complement $B$-decay measurements by providing an opportunity to master hadronic 
effects.  Heavy-quark symmetry predicts relationships between processes involving $c$ quarks and those 
involving $b$ quarks in very general ways, and specific theoretical tools like lattice QCD (LQCD) can be used 
to leverage knowledge gleaned from charm into precise predictions about $B$.

In leptonic decays the hadronic effects are packaged in the decay constant ($f_{D^+}$, $f_{D_s}$), as 
shown in Eq.~\ref{eq-LEP}.
\begin{equation}
\Gamma(D^+ \rightarrow \ell^+ \nu)=
\frac{G_F^2}{8 \pi} f_{D^+}^2 m_\ell^2 M_{D^+} \biggl[1 - \frac{m_\ell^2}{M_{D^+}^2} \biggr]^2 |V_{cd}|^2
\label{eq-LEP}
\end{equation}
Precise tests of LQCD are possible through measurements of leptonic $D^+$ and $D_s^+$ decay constants (or, even more precisely,
their ratio).  Calculations of the corresponding quantities for $B_d$ and $B_s$ decays then can be used to 
reduce the theoretical uncertainties in extracting CKM parameters from measurements of $B_d$ and $B_s$ mixing.
In exclusive semileptonic decays the hadronic effects are bundled in the form factors, as shown for
$D$ to pseudoscalar transitions in Eq.~\ref{eq-SL}.
\begin{equation}
\Gamma(D \rightarrow P \ell \nu)=
\frac{G_F^2 |V_{cq}|^2 p^3}{24 \pi^3} |f_+(q^2)|^2
\label{eq-SL}
\end{equation}

As with much of the CLEO-c charm program, most measurements of leptonic and semileptonic decays
rely on tagging one of the $D$ mesons in the pure $D \bar{D}$ state produced in $\psi(3770)$ decays.  Full reconstruction
of a $D$ meson on one side of tagged events effectively provides a ``beam'' of $D$'s of known 4-momentum on
the other side.  Depending on the cleanliness of the process being measured, the selection of tags can
be fine-tuned, with low-multiplicity ``golden modes'' for best purity and additional modes with higher
charged multiplicity and/or $\pi^0$'s for best efficiency.  Typical tagging efficiencies in CLEO-c analyses
are in the range of 15\% for $D^0$ and 10\% for $D^+$.

\begin{figure}[h]
\centering
\includegraphics[width=80mm]{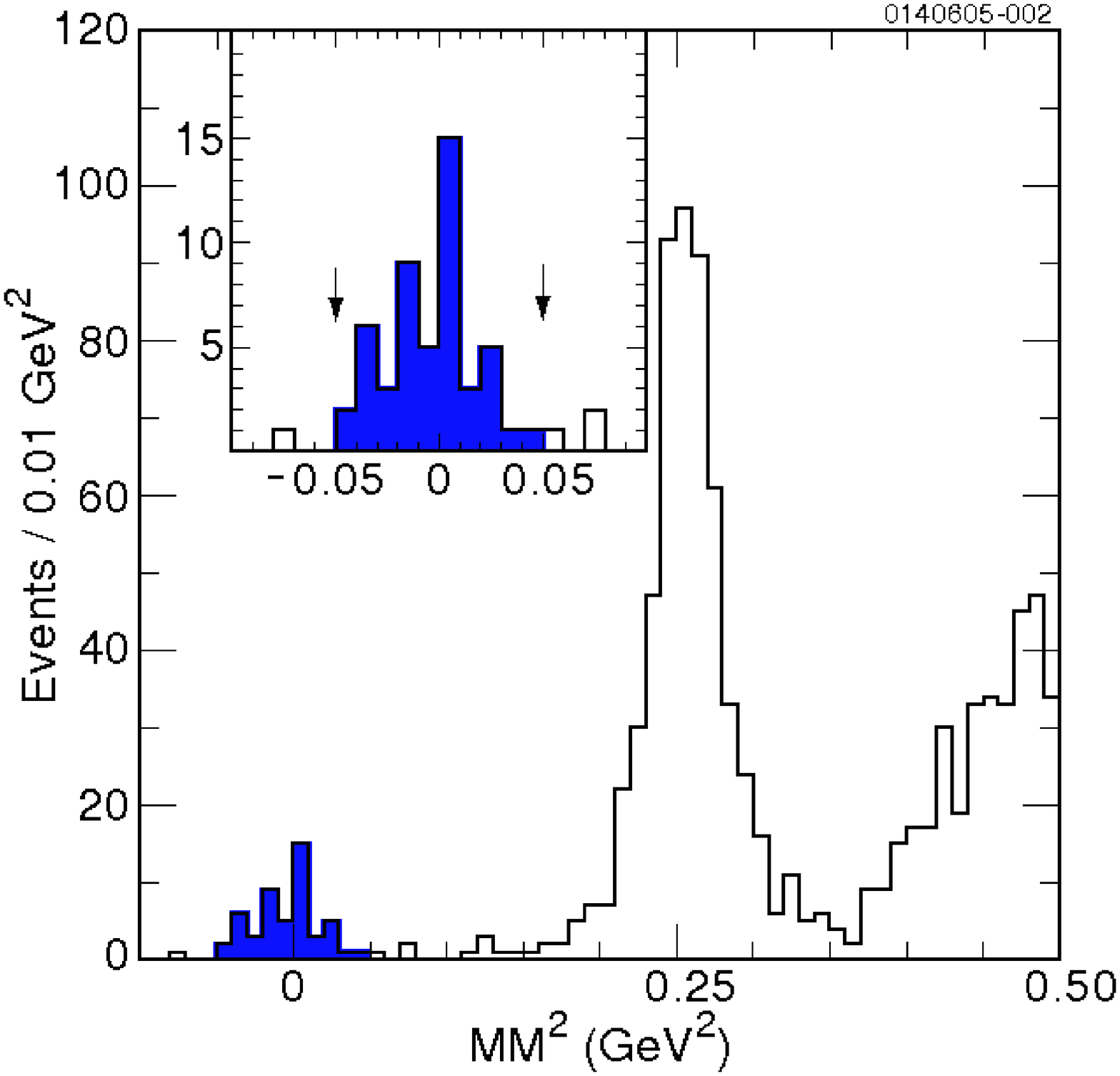}
\caption{MM$^2$ for events with selected $D^-$ tags, an oppositely-charged muon-like track,
and minimal extra energy.  The inset shows a close-up of the signal region, with cut values
indicated by the arrows.} \label{fig:Dmunu}
\end{figure}

A 2004 measurement of the leptonic decay $D^+ \rightarrow \mu^+ \nu_\mu$ was the first CLEO-c result to 
be presented, and a follow-up measurement with the current full $\psi(3770)$ data sample of 
281~$pb^{-1}$ has been published \cite{Artuso:2005ym}.  A total of about 160,000 $D^+$ tags are selected in 
six hadronic decay modes.  Signal events are required to have a tag accompanied by exactly one track of charge 
opposite to the tag, with minimal extra energy in the event.  The track is required to have a muon-like energy 
deposit of less than 300~MeV in the electromagnetic calorimeter.  True $D^+ \rightarrow \mu^+ \nu_\mu$ decays 
are selected based on the missing-mass-squared variable MM$^2$ defined in Eq.~\ref{eq-MMsq}.  
\begin{equation}
{\rm MM}^2 = (E_{beam} - E_{\mu^+})^2 - (-{\bf p}_{D^-} - {\bf p}_{\mu^+})^2
\label{eq-MMsq}
\end{equation}
The MM$^2$ distribution in the data is shown in Fig.~\ref{fig:Dmunu}. 
There is a clear signal in the vicinity of MM$^2=0$, cleanly separated from the potentially severe
background of $D^+ \rightarrow K^0_L \pi^+$.  Potential backgrounds at low MM$^2$, like 
$D^+ \rightarrow \pi^0 \pi^+$ are effectively suppressed by the extra-energy cut.  Overall, there are
50 events in the $D^+ \rightarrow \mu^+ \nu_\mu$ signal region, with a total background of 2.81, where
we use data as much as possible to estimate the larger background sources.
This leads to the following branching fraction measurement:
\begin{equation}
{\cal B}(D^+ \rightarrow \mu^+ \nu_\mu) = (4.4 \pm 0.7 \pm 0.1 \times 10^{-4}),
\label{eq-DmunuBF}
\end{equation}
and a value for the decay constant of 
\begin{equation}
f_{D^+} = (223 \pm 17 \pm 3)~{\rm MeV}.
\label{eq-DmunuDC}
\end{equation}
This result is consistent with the LQCD prediction of Aubin {\it et al.} \cite{Aubin:2005ar} of
$(201 \pm 3 \pm 17)~{\rm MeV}$.  The juxtaposition of the experimental and theoretical statistical
and systematic uncertainties illustrates that, with more data from CLEO-c, there is an excellent
opportunity to put the lattice to a rigorous test.

The CLEO-c program of leptonic-decay studies aspires to measure all accessible modes with 
the best precision possible to verify internal consistency and check for conformance
to Standard Model expectations.  In this context, Ref.~\cite{Artuso:2005ym} includes a limit on
the branching fraction for $D^+ \rightarrow e^+ \nu_e$, applying exactly the same selection
procedures except that the single candidate track is required to pass an electron-identification
requirement.  The 90\% confidence level upper limit for $D^+ \rightarrow e^+ \nu_e$ of 
$2.3 \times 10^{-5}$ is well above the Standard Model prediction, which is smaller than that for 
the muon mode by a factor of 50,000.

\begin{figure}[b]
\centering
\includegraphics[width=80mm]{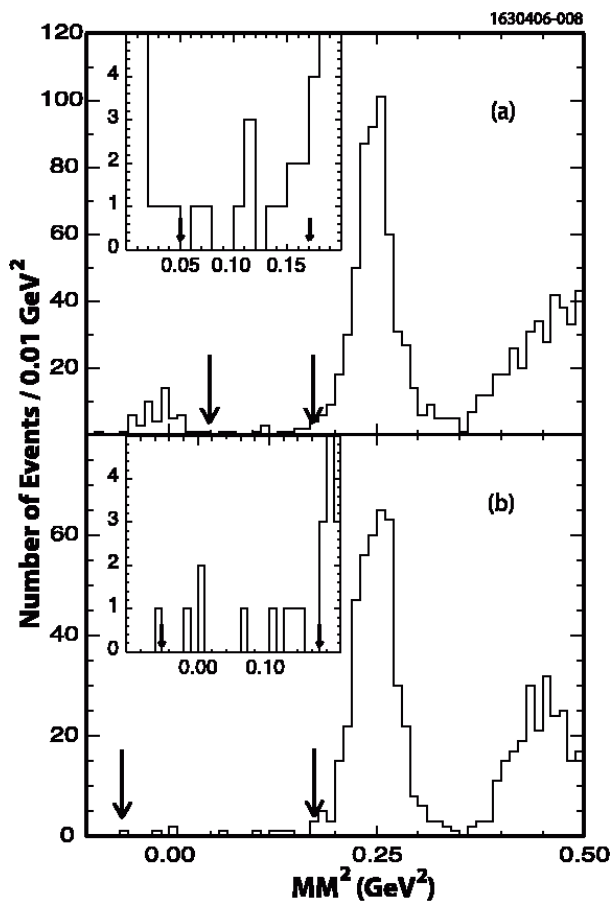}
\caption{MM$^2$ for events with selected $D^-$ tags, an oppositely-charged track with 
(a) $p<300$~MeV or (b) $p>300$~MeV, and minimal extra energy.  The limits of the
$D^+ \rightarrow \tau^+ \nu_\tau$ signal regions in the two cases are indicated by the arrows.} 
\label{fig:Dtaunu}
\end{figure}

In the same vein, CLEO-c has recently submitted to Phys. Rev.~D a search for the leptonic
decay $D^+ \rightarrow \tau^+ \nu_\tau$ \cite{Rubin:2006nt}.  This provides complementary
information to $D^+ \rightarrow \mu^+ \nu_\mu$ and potential further verification 
that leptonic charm decays present a picture consistent with the Standard Model.
The decay 
$D^+ \rightarrow \tau^+ \nu_\tau$ has a much smaller helicity suppression than 
$D^+ \rightarrow \mu^+ \nu_\mu$ because of the large mass of the $\tau$ compared to 
the $\mu$.  This is partially defeated by limited phase space, however, so that the expected 
advantage in branching fraction is only a factor of 2.65.
The limited phase space is also the reason that 
$D^+ \rightarrow \tau^+ \nu_\tau$ can be measured with a technique that is only a slight modification
of that for $D^+ \rightarrow \mu^+ \nu_\mu$, with identical tag selection and procedures
for extracting signal yields and estimating background.  The candidates are divided into 
subsamples that are either muon-like (calorimeter energy less than 300~MeV) 
or not (calorimeter energy greater than 300~MeV - rare for muons, less so for pions).  
The presence of the extra neutrino from the $\tau$ decay produces a distribution in MM$^2$, 
rather than the narrow peak of the $D^+ \rightarrow \mu^+ \nu_\mu$.  The slowness of the 
$\tau$ keeps MM$^2$ values low, however, populating the region between the 
$D^+ \rightarrow \mu^+ \nu_\mu$ and $D^+ \rightarrow K^0_L \pi^+$.  Subdividing the 
sample increases the efficiency of the selection, since for pion-like tracks we can include 
the MM$^2=0$ region in the signal.  The MM$^2$ distributions are shown in 
Fig.~\ref{fig:Dtaunu}. 
For the $p<300$~MeV ($p>300$~MeV) sample there are 12 (8) candidates with an estimated 
background of $6.1 \pm 0.6 \pm 0.3$ ($5.0 \pm 0.6 \pm 0.2$), giving a net yield of 
$5.9 \pm 3.5 \pm 0.3$ ($3.0 \pm 2.9 \pm 0.2$). In neither case is there a statistically 
significant excess, and the resulting upper limit is 
\begin{equation}
{\cal B}(D^+ \rightarrow \tau^+ \nu_\tau) < 2.1 \times 10^{-4}~(90\%~{\rm confidence}).
\label{eq-DtaunuBF}
\end{equation}
This does not yet constitute a rigorous test of the Standard Model prediction, which
is larger than this limit by a factor of 1.8.

\begin{figure}[h]
\centering
\includegraphics[width=80mm]{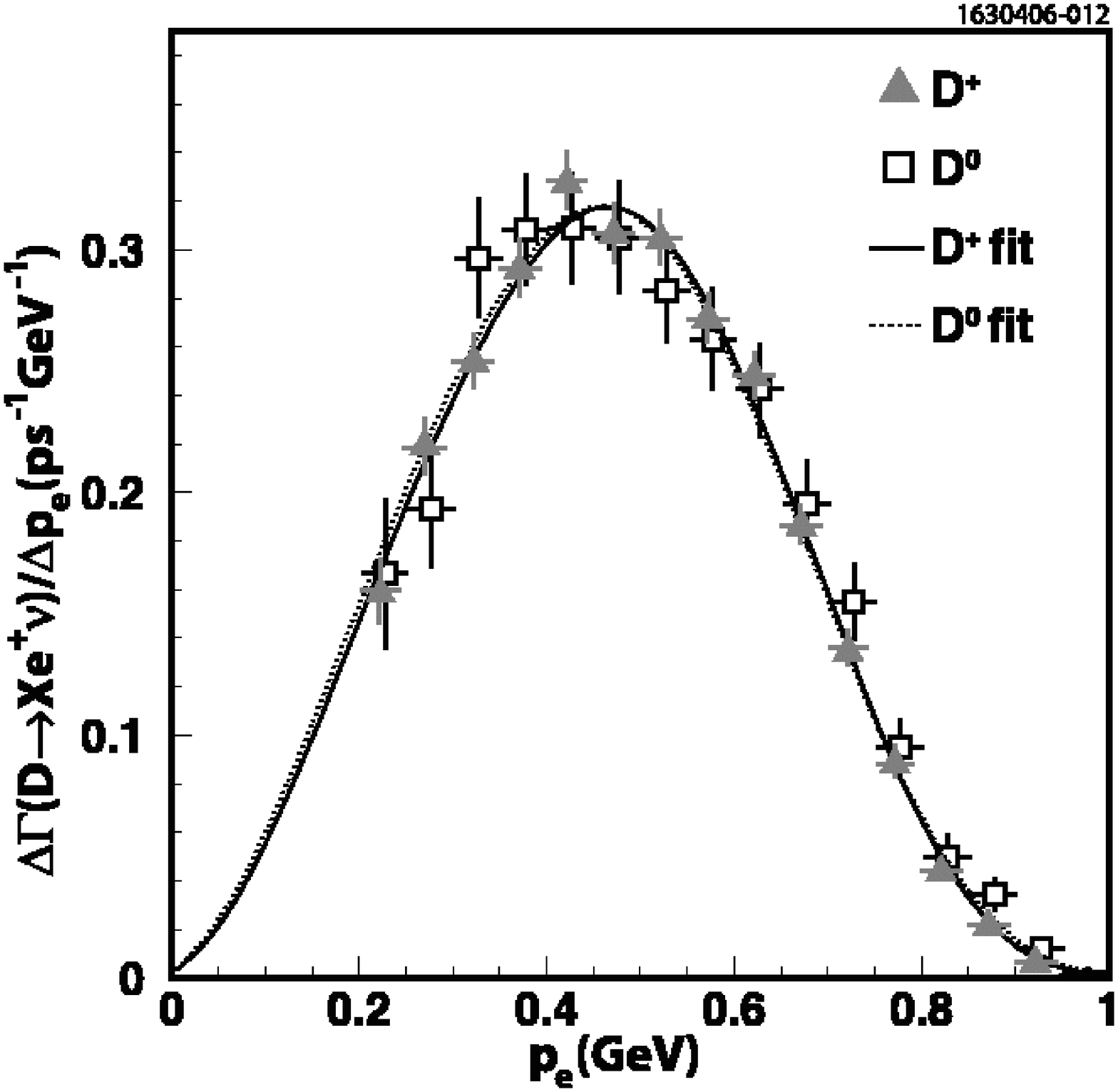}
\caption{\label{fig:SL_spec}Positron differential semileptonic widths
$d\Gamma^{\text{sl}}/dp_e$ for  the decays $D^+\rightarrow X e^+
\nu_e$ (filled triangles) and $D^0\rightarrow Xe^+\nu_e$  (open
squares) in the laboratory frame.
The errors shown include statistical errors and additive
systematic errors. The symbols for $D^+$ and $D^0$ spectra are
shifted for clarity. The curves represent 
the fits used to extrapolate the measured spectra below
200~MeV/$c$.} 
\end{figure}

In the realm of semileptonic charm decays CLEO-c has both inclusive and exclusive analyses 
presented and forthcoming that represent significant advances beyond previous knowledge. 
Inclusive decays $D \rightarrow X e^+ \nu_e$ are of interest as a cross-check of 
exclusive measurements, to determine whether the apparent saturation by the lightest
vector and pseudoscalar modes survives to higher precision or if there is significant 
room for as-yet-unobserved exclusive modes.  Comparisons of the charged and neutral 
semileptonic widths probe for contributions other than tree-level diagrams, such as the
weak-annihilation processes that constitute a possible background for charmless semileptonic
$B$ decays \cite{Rosner:2006zz}.  Finally, the inclusive measurements are important
input for other experiments, which need the branching fractions and momentum spectra for precise
modeling of secondary charm decays in $B$ studies.  CLEO-c has recently presented
measurements of inclusive $D^0$ and $D^+$ semileptonic decays based on the 281~pb$^{-1}$
$\psi(3770)$ sample \cite{Adam:2006nu}.  This analysis uses only the purest tags, roughly 
47,000 $D^0 \rightarrow K^- \pi^+$ (signal/background $\sim 60$) and 74,000 
$D^+ \rightarrow K^- \pi^+ \pi^-$ (S/B $\sim 25$).  Electrons are measured with a 
minimum-momentum requirement of 200~MeV/$c$, giving access to about 92\% of the full spectrum.
The solid-angle acceptance of 80\% of $4 \pi$ corresponds to the fiducial volume of the 
RICH detector.  The 
electron-identification efficiency, determined with data, is about 95\% above 
300~MeV/$c$ and about 71\% between 200 and 300~MeV/$c$.  The probability of misidentifying 
a charged hadron as an electron, also determined with data, is about 0.1\% over most of 
the momentum range.  We rely on charge correlations to eliminate charge-symmetric
electron backgrounds due to photon conversions and $\pi^0$ Dalitz decays.  The yield
measurement is summarized in Table~\ref{tab:DSLinc}, 
\begin{table}[htpb]
\begin{center}
\caption{\label{tab:DSLinc}  Positron unfolding procedure and
corrections for the inclusive $D$ semileptonic measurements. The errors reported in the intermediate yields reflect
only statistical uncertainties.} \vskip 0.5 cm
\begin{tabular}{lcc}
\hline ~~& $D^+$ & $D^0$ \\
\hline
Signal $e^+$& ~~ & ~~ \\
Right-sign (raw)& $8275\pm 91$ & $ 2239\pm 47$ \\
Wrong-sign (raw) & $228\pm 15$ & $233 \pm 15$ \\
\hline Right-sign (unfolded) & $9186\pm 103$ & $2453\pm 54$\\
Wrong-sign (unfolded) & $231\pm 19$ & $ 203\pm 19$ \\
Sideband $e^+$(RS) & 168$\pm$ 13 & 15 $\pm$ 4 \\
Sideband $e^+$(WS) & $11\pm 5$ & $11\pm 4$ \\
Net $e^+$ & $8798 \pm 105$ & $2246\pm 57$ \\
Corrected Net $e^+$ & $10998\pm 132$ & $2827\pm 72$\\\hline
\end{tabular}
\end{center}
\end{table}
and the lab-frame momentum spectra for semileptonic $D^+$ and $D^0$ 
decays are shown in Fig.~\ref{fig:SL_spec}.
$D$-frame spectra will be presented in a forthcoming publication.
There is good agreement between the spectra for charged and neutral
$D$'s, and both are fitted to theoretical models that include 
final-state radiation to determine the correction for the 
unmeasured portion of the spectra.  The resulting branching 
fraction measurements are given in Eqs.~\ref{eq-DSLinclBFpl} and 
\ref{eq-DSLinclBF0}.
\begin{equation}
{\cal B}(D^{+}\rightarrow Xe^+\nu _e) = (16.13 \pm 0.20_{\rm stat} \pm
   0.33_{\rm sys})\%
\label{eq-DSLinclBFpl}
\end{equation}
\begin{equation}
{\cal B}(D^{0}\rightarrow Xe^+\nu _e) =  (6.46 \pm 0.17_{\rm stat}  \pm
   0.13_{\rm sys})\%
\label{eq-DSLinclBF0}
\end{equation}
Combining these branching fractions with measurements of $D$ lifetimes
\cite{pdg04} gives values for the total inclusive semileptonic widths of
$\Gamma(D^{+}\rightarrow Xe^+\nu _e) = 0.1551 \pm 0.0020 \pm 0.0031$
ps$^{-1}$ and $\Gamma(D^{0}\rightarrow Xe^+\nu _e) = 0.1574 \pm
0.0041 \pm 0.0032$ ps$^{-1}$. Their ratio, 
$\Gamma_{D^+}^{\text{sl}}/\Gamma_{D^0}^{\text{sl}}=0.985\pm 0.028\pm
0.015$, is consistent with the expectation of isospin invariance.

CLEO-c has previously published measurements of the branching fractions
for five $D^+$ \cite{Huang:2005iv} and five $D^0$ \cite{Coan:2005iu} 
exclusive semileptonic modes based on the first 57~pb$^{-1}$ of $\psi(3770)$ 
data. In every case these are improvements over previous world averages
\cite{pdg04}, and two of the modes ($D^+ \rightarrow \omega e^+ \nu_e$
and $D^0 \rightarrow \rho^- e^+ \nu_e$) are first measurements.  Extensions
of this exclusive semileptonic study to 281~pb$^{-1}$, including additional 
modes and form-factor measurements, are in progress.  Results are expected in
summer 2006.

A new preliminary CLEO-c analysis takes an alternative approach to 
measuring the branching fractions and form factors
for the exclusive decays $D \rightarrow P e \nu$, where $P$ signifies a
pseudoscalar ($\pi$ or $K$).  In this case 
there is no $D$-tag requirement, but rather full reconstruction of the 
semileptonic decay from all decay products, including the neutrino.  The
technology of ``neutrino reconstruction'' was developed for CLEO~II initially
for exclusive $b \rightarrow u \ell \nu$ measurements.  The measured missing 
momentum and energy in the event are assigned to the neutrino:
$P_\nu \equiv P_{\rm event} - P_{\rm visible}$, allowing computation
of the event $\Delta E$, $M_{bc}$, and $q^2$.  
Special track- and shower-selection criteria ensure, to the maximum extent 
possible, that all true particles are included, with
no false ones or duplicates.  To make this feasible, events are chosen  
selectively: no net charge and only one identified lepton, since
a second lepton signals the presence of a second neutrino.

This procedure yields one-sigma resolution in the core of the measured
$P_{\rm miss}$ distribution of $\sim 13$~MeV.  The resolution in the missing energy
is significantly poorer, so the measured missing energy is replaced by
the magnitude of the missing momentum.  In computing $M_{bc}$ and
$q^2$, the resolution is improved by scaling the 4-momentum to 
satisfy $\Delta E$=0.

\begin{figure*}[t]
\centering
\includegraphics[width=135mm]{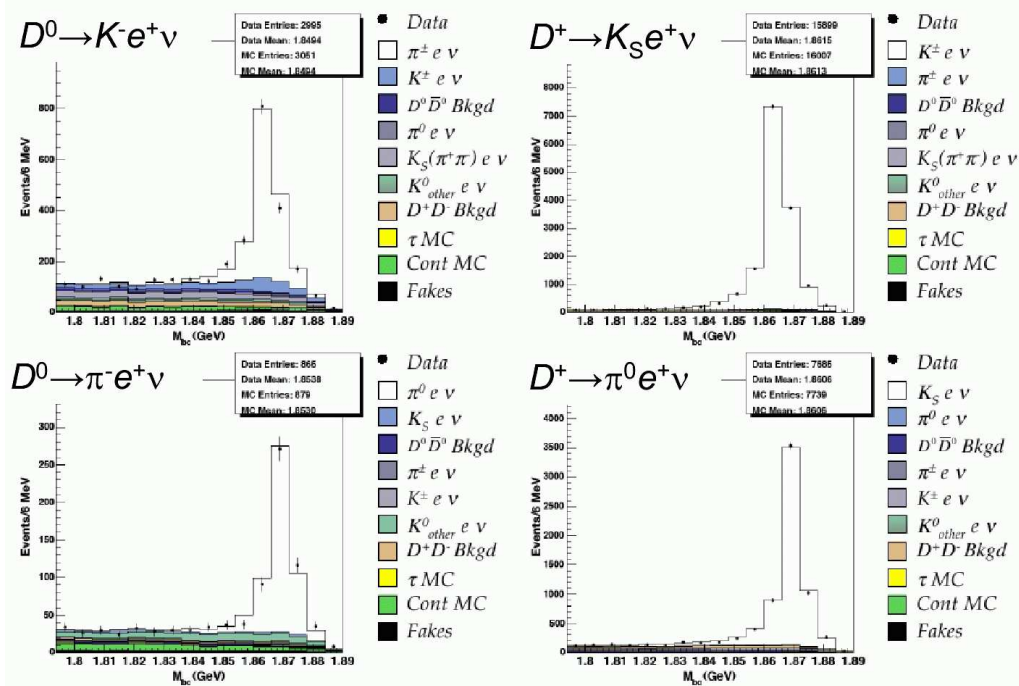}
\caption{\label{Excl_NR_fits}Fits to $M_{bc}$ distributions for
all four $D \rightarrow P e \nu$ modes summed over all $q^2$ bins.} 
\end{figure*}

Semileptonic decay candidates are selected by cutting on $\Delta E$ and
fitting $M_{bc}$ distributions.  This is done in five bins in $q^2$ and
the resulting $d{\cal B}/dq^2$ distributions can be integrated to get the
branching fractions and fitted to obtain the form-factor parameters.
$M_{bc}$ fits summed over all $q^2$ bins for all four $D \rightarrow P e \nu$
modes are shown in Fig.~\ref{Excl_NR_fits}.
The signals are clear and the fits, which include backgrounds from data
and simulations, are excellent.  The resulting branching fractions are as
follows:
\begin{equation}
{\cal B}(D^{0}\rightarrow K^- e^+ \nu _e) =  (3.56 \pm 0.03 \pm
   0.10)\%
\label{eq-Excl_NR_Km}
\end{equation}
\begin{equation}
{\cal B}(D^{0}\rightarrow \pi^- e^+ \nu _e) =  (0.301 \pm 0.011 \pm
   0.010)\%
\label{eq-Excl_NR_pim}
\end{equation}
\begin{equation}
{\cal B}(D^{+}\rightarrow K^0 e^+ \nu _e) =  (8.70 \pm 0.13 \pm
   0.27)\%
\label{eq-Excl_NR_Kz}
\end{equation}
\begin{equation}
{\cal B}(D^{+}\rightarrow \pi^0 e^+ \nu _e) =  (0.381 \pm 0.025 \pm
   0.015)\%
\label{eq-Excl_NR_piz}
\end{equation}
The ratios of branching fractions and partial widths
are found to be ${\cal B}(D^0 \to \pi^- e^+ \nu) \slash {\cal B}(D^0 \to
K^- e^+ \nu) = 0.085 \pm 0.003 \pm 0.001$, ${\cal B}(D^+ \to \pi^0 e^+
\nu) \slash {\cal B}(D^+ \to \bar{K}^0 e^+ \nu) = 0.044 \pm 0.003 \pm 0.001$,
$\Gamma(D^0 \to \pi^- e^+ \nu) \slash \Gamma(D^+ \to \pi^0 e^+ \nu) =
2.04 \pm 0.15 \pm 0.08$, and $\Gamma(D^0 \to K^- e^+ \nu) \slash
\Gamma(D^+ \to \bar{K}^0 e^+ \nu) = 1.06 \pm 0.02 \pm 0.03$.  The 
partial-width ratios are consistent with isospin symmetry.

The results of the fits for the form-factor parameters 
to the measured $q^2$ distributions with the 
simple pole \cite{richman} and the modified pole \cite{BKparam}
models are given in Table~\ref{table:all_values}. 
\begin{table}[hptb]
\begin{center}
\begin{tabular}{ccc}
\hline
\hline
\multicolumn{3}{c}{{\it Simple Pole Model }} \\
{\it Decay Mode} & $|V_{cx}|f^+(0)$ & $m_{\mathrm{pole}}$ \\
\hline
$D^0 \to \pi^- e^+ \nu$ & $0.146 \pm 0.004 \pm 0.003$ & $1.87 \pm 0.03
\pm 0.01$  \\
$D^0 \to K^- e^+ \nu$ & $0.736 \pm 0.005 \pm 0.010$ & $1.98 \pm 0.03
\pm 0.02$  \\
$D^+ \to \pi^0 e^+ \nu$ & $0.152 \pm 0.007 \pm 0.004$ & $1.97 \pm 0.07
\pm 0.02$  \\
$D^+ \to \bar{K}^0 e^+ \nu$ & $0.719 \pm 0.009 \pm 0.012$ & $1.97 \pm 0.05
\pm 0.02$  \\
\hline
\hline
\multicolumn{3}{c}{{\it Modified Pole Model }} \\
{\it Decay Mode} & $|V_{cx}|f^+(0)$ & $\alpha$ \\
\hline
$D^0 \to \pi^- e^+ \nu$ & $0.142 \pm 0.005 \pm 0.003$ & $0.37 \pm 0.09
\pm 0.03$  \\
$D^0 \to K^- e^+ \nu$ & $0.734 \pm 0.006 \pm 0.010$ & $0.19 \pm 0.05
\pm 0.03$  \\
$D^+ \to \pi^0 e^+ \nu$ & $0.151 \pm 0.008 \pm 0.004$ & $0.12 \pm 0.17
\pm 0.05$  \\
$D^+ \to \bar{K}^0 e^+ \nu$ & $0.718 \pm 0.009 \pm 0.012$ & $0.20 \pm 0.08
\pm 0.04$  \\
\hline
\end{tabular}
\caption{ {Simple-pole model and modified-pole model form-factor parameters resulting 
from fits to the $d{\cal B}/dq^2$ distributions for the four $D \rightarrow P \ell \nu$
decay modes. } }
\label{table:all_values}
\end{center}
\end{table}
The $q^2$ fits yield values of the intercepts at $q^2$=0 which are
equal to $|V_{cx}|f(0)$.  These can be combined with LQCD calculations
of $f(0)$ \cite{Aubin:2004ej} to obtain preliminary results for
the magnitudes of the CKM matrix elements which are consistent with
current world averages, but are limited in precision by $\sim 10$\% theoretical 
uncertainties.

The precision of the semileptonic measurements made by the 
neutrino-reconstruction is competitive with the tagged results.  Because
the techniques are so different, there is considerable statistical
independence and the best CLEO-c results will come from averaging the
two techniques.

\section{The CLEO-c \boldmath$D_s$ Scan}
\label{sec:scan}

When the CLEO-c program was proposed \cite{Briere:2001rn} its centerpiece was charm:
direct and indirect CKM measurements, theoretical tests, and branching fractions and 
charm properties as engineering input for other experiments.  Within the charm
program there were two components: $D^0$ and $D^+$ leptonic, semileptonic and
hadronic decays at the $\psi(3770)$ resonance, and $D_s$ leptonic, semileptonic and
hadronic decays at some energy to be determined.  The mechanism for determining
that energy was a two-month scan run that was carried out in late summer and early 
fall of 2005.  

\begin{figure}[b]
\centering
\includegraphics[width=80mm]{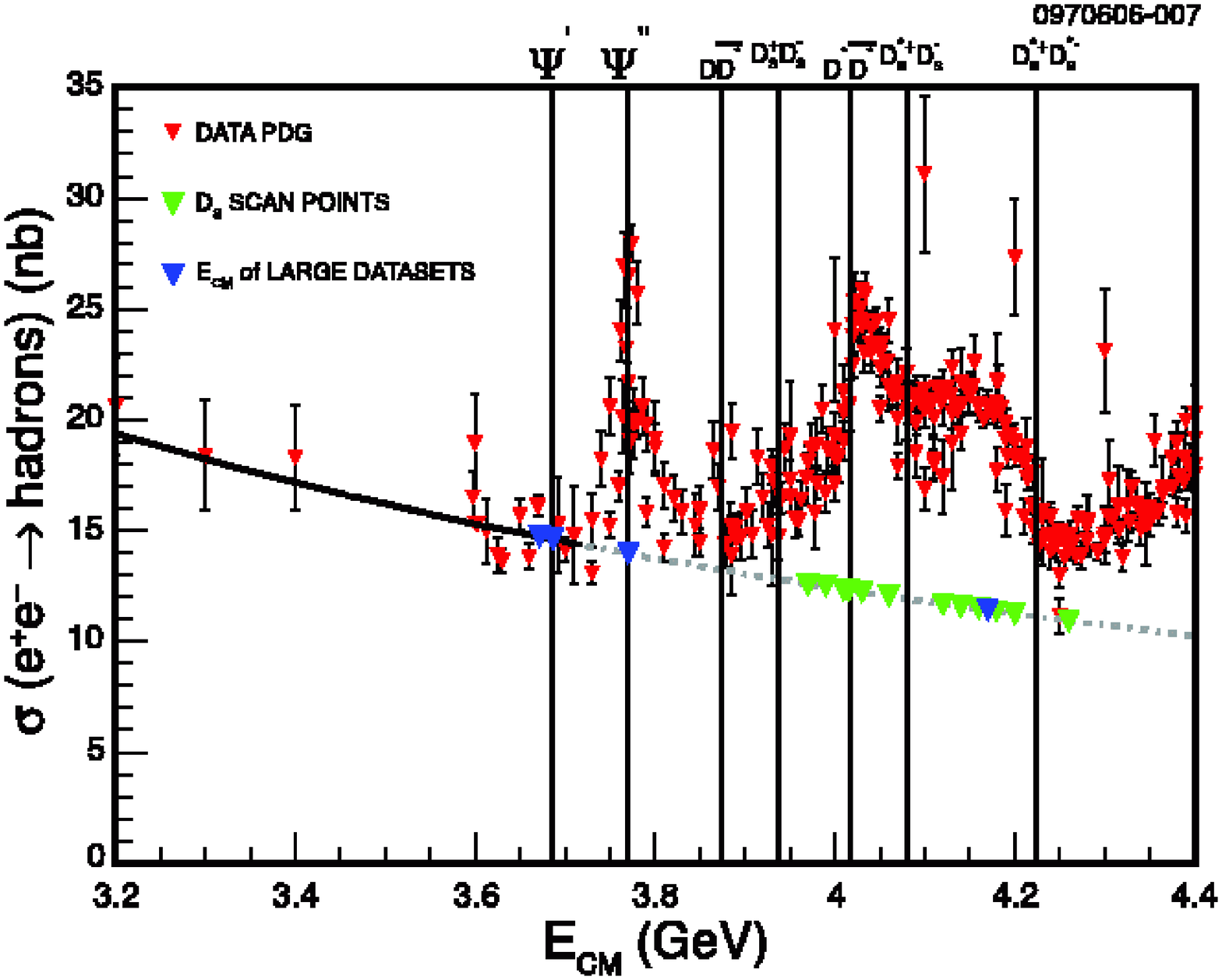}
\caption{\label{fig:Scan_Rplot}Data on hadron production ($R$) as a function
of $e^+e^-$ center-of-mass energy.  Inverted green triangles
indicate energies of the CLEO-c scan-run points and inverted
blue triangles denote larger data samples at $\psi(2S)$, $\psi(3770)$, and 
$E_{CM}=4170$~MeV, the last collected after the scan run.  
Prominent charmonium states and threshold energies 
for open-charm final states are indicated.} 
\end{figure}

The principal goal of the scan was determination of the optimal energy for
studying $D_s$ physics.  That optimum depends both on the production cross section 
and the complexity of the events that dominate $D_s$ production.  This had to be
measured by CLEO-c because, while there is beautiful data on hadron production
in the region of charm threshold (Fig.~\ref{fig:Scan_Rplot}) \cite{pdg04}, there 
is little experimental detail on its composition and theoretical guidance 
is limited.  Secondary motivations for the scan included assessing our ability
to do $D^+$ and $D^0$ physics at energies above $\psi(3770)$, disentangling the
``intricate behavior'' of charm production in the region above charm threshold
\cite{Voloshin:2006pz}, and confirming and investigating the
$Y(4260)$ state discovered by BaBar last year \cite{Aubert:2005rm}.

\begin{figure}[h]
\centering
\includegraphics[width=80mm]{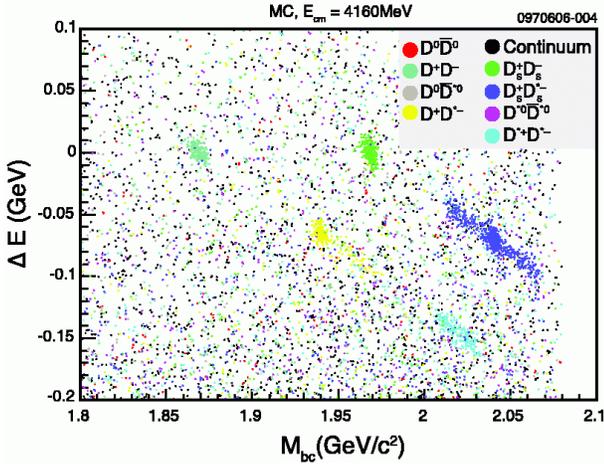}
\caption{\label{fig:Scan_Ds2D}$\Delta E$ vs. $M_{bc}$ for selected 
$D_s^+ \rightarrow \phi \pi^+$ candidates for Monte Carlo generated at 
$E_{CM}=4160$~MeV.  There is clear separation of candidates into different 
event types.} 
\end{figure}

The scan that was conducted consisted of twelve points in the energy region 3970-4260~MeV, 
totaling about 60~pb$^{-1}$.  At each energy point the first task was to make quick 
determinations of the cross sections for each of the two-charmed-meson final states 
($D^{(*)}_{(s)}{\bar D}^{(*)}_{(s)}$) that were accessible at that energy.  
This can be done cleanly and efficiently without
attempting full reconstruction of $D^*$ states by selecting $D$ and $D_s$ mesons
with the usual tools of $D$ tagging.  Three, five and eight decay modes were employed for
$D^0$, $D^+$ and $D^+_s$, respectively.  The different production channels can
be distinguished based on the kinematics of reconstructed tags reflected through the
selection variables $\Delta E$ and $M_{bc}$, the latter of which is equivalent to 
$D_{(s)}$ momentum.  Fig.~\ref{fig:Scan_Ds2D} shows a 
two-dimensional plot of $\Delta E$ vs. M$_{bc}$ for
$D_s \rightarrow \phi \pi$ candidates from a Monte
Carlo sample generated at 4160~MeV.  There are clear concentrations of events
at characteristic values. The $D_s^+D_s^-$ appears as a tight cluster at $\Delta E$=0 
and $M_{bc}$=$M_{D_s}$, for example, while the $D_s^+D_s^{*-}$ is an elliptical
structure with a central core consisting of direct $D_s$ and smeared wings
with $D_s$ from $D_s^*$.  Because of Cabibbo-suppressed modes
that can populate this final state, there are also clusters for non-strange charmed-meson
final states: $D{\bar D}$, $D{\bar D}^*$ and $D^*{\bar D}^*$.

The yields for $D {\bar D}$  and $D_s {\bar D_s}$ were extracted by cutting on 
$\Delta E$ and projecting $M_{bc}$ distributions.  For the other event types the
procedure was to cut on $M_{bc}$ and project onto invariant mass to extract the yields.
All cut values were determined by kinematics, with no double counting allowed.  Cross-feed 
among modes is small and easily determined.

One complication that quickly became apparent from recoil-mass and charmed-meson momentum
distributions is that two-body modes do not account for all charmed-meson
production.  There is clear evidence for ``multibody'' production such as
$e^+ e^- \rightarrow D {\bar D}^* \pi$.  Work to thoroughly disentangle and account
for this production mechanism is under way.

\begin{figure}[h]
\centering
\includegraphics[width=80mm]{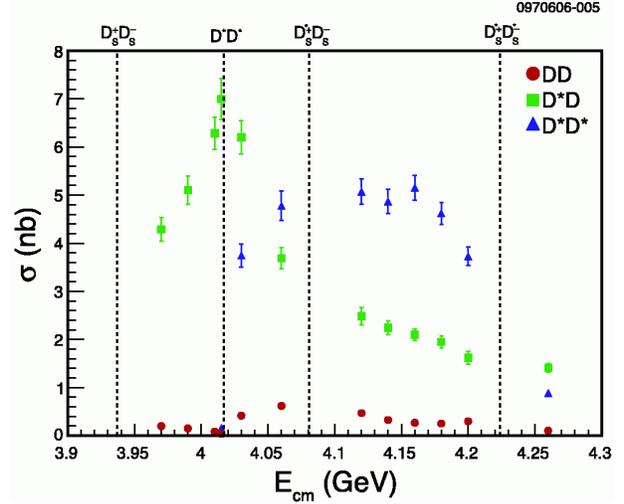}
\caption{\label{fig:Scan_DD}Cross sections for $D {\bar D}$, $D {\bar D}^*$, 
and $D^* {\bar D}^*$ from the CLEO-c scan run.} 
\end{figure}

Preliminary cross-section results (partially evaluated systematic uncertainties, no 
correction for multibody contributions, not radiatively corrected) for the non-strange 
charmed meson combinations are shown in Fig.~\ref{fig:Scan_DD}.  There is very little 
$D {\bar D}$ production
at any energy, a sharp peak structure in $D {\bar D}^*$ near $D^* {\bar D}^*$
threshold, and a broad peak or plateau in $D^* {\bar D}^*$. The charm cross section 
through this region is considerable, comparable to that at the $\psi(3770)$. 

\begin{figure}[h]
\centering
\includegraphics[width=80mm]{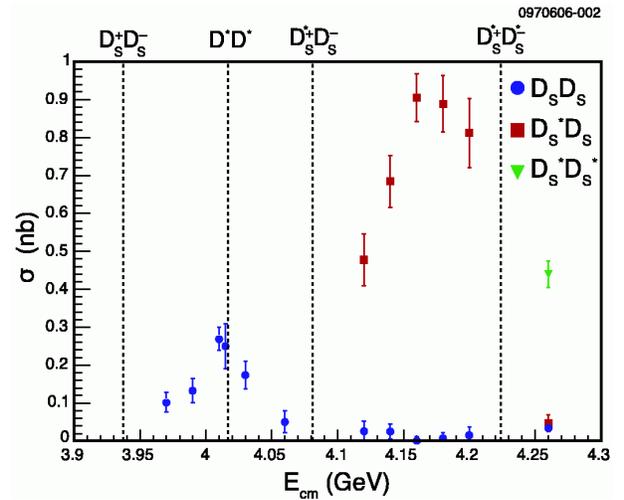}
\caption{\label{fig:Scan_Ds}Cross sections for $D_s {\bar D}_s$, $D_s {\bar D}_s^*$, 
and $D^*_s {\bar D_s}^*$ (only accessible at the highest point) from the CLEO-c scan run.} 
\end{figure}

The corresponding (preliminary) bottom line for $D_s$ production is 
shown in Fig.~\ref{fig:Scan_Ds}.
In this case there is a visible, but disappointingly small, peak in 
$D_s {\bar D}_s$, and a more impressive broad peak at about 4170~MeV which
offers about 1~nb of $D_s$ production.

\begin{figure}[h]
\centering
\includegraphics[width=80mm]{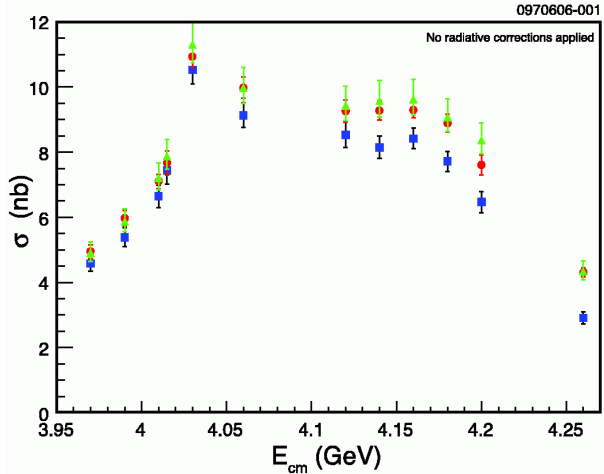}
\caption{\label{fig:Scan_IncExc}Cross sections for (blue) exclusive-charm measurement obtained by 
summing two-charmed-meson final states and two inclusive measurements: (red)
straightforward inclusive counting of $D^0$, $D^+$ and $D_s$ cross sections, and (green)
the total inclusive hadronic cross section minus the extrapolated $uds$ continuum.} 
\end{figure}

Many cross-checks of these results are being carried out.  A straightforward one
(Fig.~\ref{fig:Scan_IncExc}) is a comparison of the exclusive measurements described 
so far with two inclusive measurements of charm production.  One of these is the sum of 
the inclusive cross sections for $D^0$, $D^+$, and $D_s^+$ production and the second is a 
measurement of the total hadronic cross section with the extrapolated $uds$ continuum
subtracted based on lower-energy measurements. 
There is very good agreement between the two inclusive measurements, which exceed the
exclusive by an amount that grows with energy.  This is reaffirmation of the presence in
the inclusive of something that is not (effectively) counted in the exclusive: multibody
production.

Overall the scan was a great success for its primary purpose.  It succeeded in identifying
a $D_s$ ``mini-factory'' at $E_{CM}$=4170~MeV which has already been used to acquire a
sizable $D_s$ sample.  The roughly 1-nb cross-section measurement
is dependent on the input branching fractions for the eight $D_s$ selection modes.  
Refinement awaits the finalized $D_s$ branching fractions from application of the
single-tag/double-tag technique to CLEO-c 4170-MeV data.  

The successful and early achievement of the core goals of the scan left us running time 
for two additions to the original program.
The first was more running in the $D_s$ peak-production region, providing a head start 
on the accumulation of a CLEO-c 4170-MeV sample which already amounts to about 200~pb$^{-1}$, 
and which is expected to grow to about twice that in summer 2006.  The second addition was 
13.2~pb$^{-1}$ at $E_{cm} = 4260$~MeV, primarily to investigate the $Y(4260)$.

\begin{figure}[h]
\centering
\includegraphics[width=80mm]{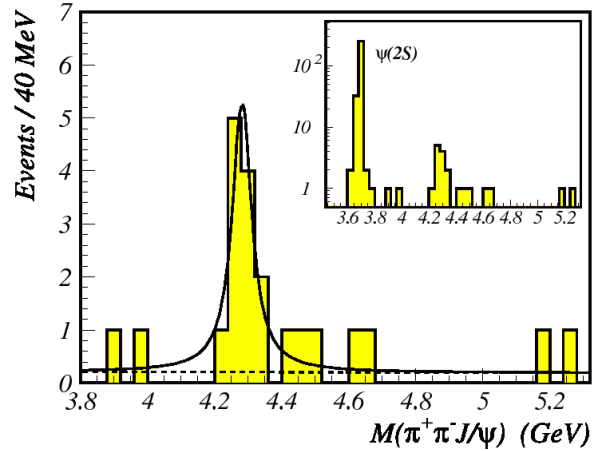}
\caption{\label{fig:CLEOIII_4260}Fit to the $\pi^+ \pi^- J/\psi$ mass distribution
(with $J/\psi \rightarrow e^+e^-$, or $\mu^+ \mu^-$) in CLEO~III ISR events.  The dotted line represents 
the fitted background and the solid curve the total fit.  The inset (semilog)
shows the same mass distribution over an extended range, allowing comparison 
with $\psi(2S)$.} 
\end{figure}

BaBar's discovery of $Y(4260)$ \cite{Aubert:2005rm} was based on a sample of 233~fb$^{-1}$
collected at the $\Upsilon(4S)$.  The observation of an enhancement in initial-state
radiation (ISR) events of the form $e^+e^- \rightarrow \gamma(\pi^+ \pi^- J/\psi)$ with effective 
energy 4260~MeV was unexpected, especially since that energy corresponds to 
a minimum in the cross section for $e^+e^- \rightarrow {\rm hadrons}$.  Independent evidence
supporting this discovery is accumulating, and there is now a preliminary observation of 
exactly the same mechanism in CLEO~III data.  The CLEO~III 
observation (Fig.~\ref{fig:CLEOIII_4260}) is based on 13.3~fb$^{-1}$ collected in the 
$\Upsilon(1S)$ - $\Upsilon(4S)$ energy range.  The statistical significance is 4.9$\sigma$,
and the measurements of energy of ($4283 ^{+17}_{-16} \pm 5$)~MeV and width
of ($70^{+40}_{-25} \pm 4$)~MeV are consistent with BaBar, although less precise.

\begin{figure}[h]
\centering
\includegraphics[width=75mm]{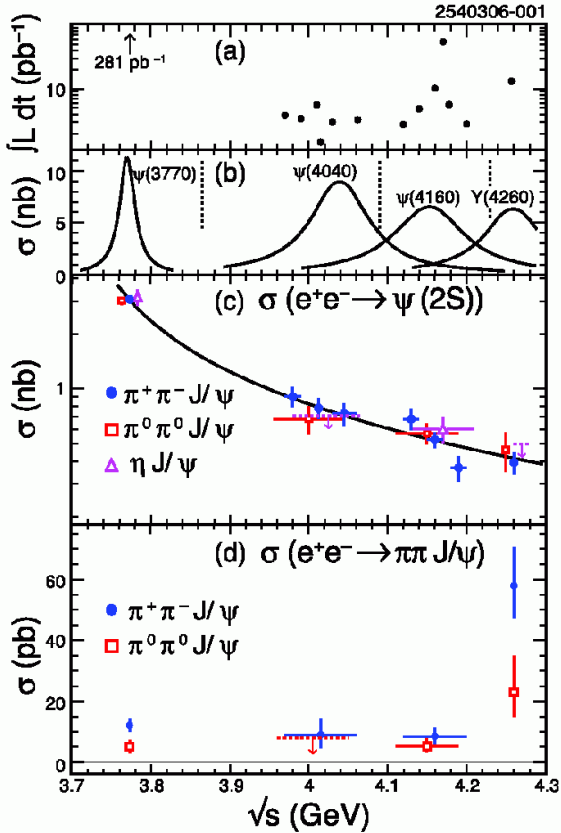}
\caption{\label{fig:CLEOc_4260}(a) Integrated luminosity vs. center-of-mass energy 
for the scan.  (b) Born-level Breit-Wigner cross sections for recognized charmonium
states (true cross sections) and $Y(4260)$ (arbitrary scale).
(c) $e^+e^- \rightarrow \gamma \psi(2S)$ cross-section measurement vs. energy for 
$\pi^+ \pi^- J/\psi$ (circles), $\pi^0 \pi^0 J/\psi$ (squares, dashed lines)
and $\eta J/\psi$ (triangles) overlaid with theoretical prediction.  (d) Cross
sections for $e^+ e^- \rightarrow \pi^+ \pi^- J/\psi$ (circles) and 
$e^+ e^- \rightarrow \pi^0 \pi^0 J/\psi$ (squares, dashed lines) vs.
energy.  Some points in (c) and (d) are offset by 10~MeV for clarity.} 
\end{figure}

There is a wide range of proposed explanations of the $Y(4260)$, including some that
are highly speculative.  They include hybrids, tetraquarks, molecular states, baryonium,
and conventional charmonium with quantum mechanical complications.  Clearly more information
was needed, and since the original observation in ISR demonstrates an assignment of 
$J^{PC} = 1^{--}$, a search for direct production in $e^+ e^-$ annihilation by CLEO-c
was a natural next step.  Readiness for this study was excellent, since this process is
closely related to others that have already been investigated in $\psi(2S)$ and
$\psi(3770)$ data.  The CLEO-c analysis \cite{Coan:2006rv} is a search for $Y(4260)$ in
sixteen possible final states with $J/\psi$, $\psi(2S)$, $\chi_{cJ}$, and $\phi$.
It is summarized in Fig.~\ref{fig:CLEOc_4260}.
This analysis benefits from the availability of the ISR process 
$e^+ e^- \rightarrow \gamma \psi(2S)$, which allows verification of the
efficiencies, background estimates and integrated luminosity for
the measurement (Fig.~\ref{fig:CLEOc_4260}(c)).  Fig.~\ref{fig:CLEOc_4260}(d)
shows that the $\psi(3770)$ and the energy regions of the $\psi(4040)$ and
$\psi(4160)$ exhibit small and consistent levels of $\pi \pi J/\psi$, while there
is strong evidence of excess production at 4260~MeV.  The data provide confirmation
($11 \sigma$) of $Y(4260) \rightarrow \pi^+ \pi^- J/\psi$, a first observation
($5.1 \sigma$) of $Y(4260) \rightarrow \pi^0 \pi^0 J/\psi$ at about half the
charged rate, and first evidence ($3.7 \sigma$) for 
$Y(4260) \rightarrow K^+ K^- J/\psi$.  These results tend to disfavor conventional 
charmonium interpretations \cite{Llanes-Estrada:2005hz} and $\chi_c \rho^0$
molecular states \cite{Liu:2005ay}, while being compatible with hybrid charmonium and 
tetraquark interpretations.  Studies of open-charm production could help clarify
the situation, but it is questionable if the 13.2~pb$^{-1}$ CLEO-c sample will be 
sufficient.

\section{Conclusion}

CLEO-c is having a major impact on the world's quantitative understanding
of charm decays, with broad implications for all of flavor physics.
Measurements of leptonic, semileptonic, and hadronic $D$ decays are already
more precise than previous PDG values, based on a small portion of the expected
data sample.   Results from the CLEO-c $D_s$ scan, including
measurements of $Y(4260)$, are bonuses beyond the original project plan.
CLEO-c is scheduled to run through April 2008, with expected
integrated luminosity that, while short of design, will still be
sufficient to realize most of the program's goals: $\sim 750$~pb$^{-1}$ each at 
$\psi(3770)$ for $D$ physics and at 4170~MeV for $D_s$ (and $D$) physics.
In addition, there will be a sample of at least 30 million $\psi(2S)$ for
charmonium physics.  The final products will include branching fractions
for $D^0 \rightarrow K^- \pi^+$,  $D^+ \rightarrow K^- \pi^+ \pi^+$, and
$D_s^+ \rightarrow \phi \pi^+$ with precisions of 
$\pm 1.25 \%$, $\pm 1.4 \%$, and 
$\pm 4 \%$, respectively.  The ultimate precision for both $D$ and $D_s$
leptonic-decay branching fractions will be less than 9\%.  These 
decay-constant measurements, detailed results on semileptonic decays, including 
form factors, and other high-precision measurements should provide the 
promised and much-needed tests of LQCD and other implementations of 
strong-interaction theory.

\bigskip 
\begin{acknowledgments}
I gratefully acknowledge the CESR accelerator staff for their continuing efforts 
to provide CLEO-c with excellent luminosity and running conditions.  CLEO-c 
enjoys generous support from the U.S. National Science Foundation and 
Department of Energy, and the Natural Sciences and Engineering Research Council of 
Canada.  Personal communications with numerous experimental and theoretical colleagues,
especially M. Voloshin, are much appreciated.
\end{acknowledgments}

\bigskip 

\end{document}